\documentclass[pdflatex,sn-mathphys-num]{sn-jnl}
\usepackage{graphicx}%
\usepackage{multirow}%
\usepackage{amsmath,amssymb,amsfonts}%
\usepackage{amsthm}%
\usepackage{mathrsfs}%
\usepackage[title]{appendix}%
\usepackage{textcomp}%
\usepackage{manyfoot}%
\usepackage{booktabs}%
\usepackage{algorithm}%
\usepackage{algorithmicx}%
\usepackage{algpseudocode}%
\usepackage{listings}%

\usepackage{mathtools}
\usepackage{braket}
\usepackage{slashed} 
\usepackage{physics}
\usepackage{bbding}
\usepackage{bbold}

\usepackage{hyperref}

\theoremstyle{thmstyleone}%

\theoremstyle{thmstyletwo}%

\theoremstyle{thmstylethree}%

\newcommand{\be}{\begin{equation}}
\newcommand{\ee}{\end{equation}}
\newcommand{\bea}{\begin{eqnarray}}
\newcommand{\eea}{\end{eqnarray}}

\newcommand{\bc}{\begin{center}}
\newcommand{\ec}{\end{center}}

\newcommand{\hk}{{\bf \hat{k}}}
\newcommand{\hr}{{\bf \hat{r}}}
\newcommand{\hn}{{\bf \hat{n}}}
\newcommand{\hp}{{\bf \hat{p}}}

\newcommand{\Bell}{{\cal B}}
\newcommand{\II}{{\cal I}_3}

\newcommand{\A}{\scriptscriptstyle{A}}
\newcommand{\B}{\scriptscriptstyle{B}}

\newcommand{\cmb}{\mathscr{C}_2}
\newcommand{\CC}{\operatorname{C}}
\newcommand{\BB}{\operatorname{B}}

\def\di{\mbox{d}}

\def\eq#1{{Eq.~(\ref{#1})}}

\begin{document}

\title[Article Title]{Quantum Entanglement and Bell Nonlocality
at Future Lepton Colliders}


\author*[1,2,3]{\fnm{Emidio} \sur{Gabrielli}}\email{emidio.gabrielli@cern.ch}
%\author[3,4]{\fnm{Luca} \sur{Marzola}}\email{luca.marzola@cern.ch}

\author[3,4]{\fnm{Luca} \sur{Marzola}}\email{luca.marzola@cern.ch}
\equalcont{This author contributed equally to this work.}

\affil[1]{\orgdiv{Department of Physics}, \orgname{University of Trieste}, \orgaddress{\street{Strada Costiera 11}, \city{Trieste}, \postcode{I-34151}, \country{Italy}}}

\affil[2]{ \orgname{INFN, Sezione di Trieste}, \orgaddress{\street{Via Valerio 2}, \city{Trieste}, \postcode{I-34127}, \country{Trieste}}}

\affil[3]{\orgdiv{Laboratory of High-Energy and Computational Physics}, \orgname{NICPB}, \orgaddress{\street{Rav\"ala pst 10}, \city{Tallinn}, \postcode{10143}, \country{Estonia}}}

\affil[4]{\orgdiv{Institute of Computer Science}, \orgname{University of Tartu}, \orgaddress{\street{Narva mnt 18}, \city{Tartu}, \postcode{51009}, \country{Estonia}}}


\abstract{Quantum entanglement and Bell nonlocality--cornerstones of quantum mechanics--have traditionally been investigated only in low-energy experimental settings. Only recently, these fundamental phenomena have come to be explored in the high-energy domain of particle physics, where collider experiments offer a powerful new platform for studying the phenomenology of quantum correlations. We present here recent results on the detection of entanglement and Bell nonlocality in processes such as tau-lepton, $WW$, and $ZZ$ pair production,  illustrating the potential of Future Lepton Colliders to probe the quantum properties of fundamental interactions.}

%%================================%%
%% Sample for structured abstract %%
%%================================%%




\maketitle

\newpage

\section{Introduction}\label{sec1}
Quantum entanglement is among the most fascinating features of quantum mechanics. Entangled systems cannot be described with combinations of the states that describe their components, and this may hold true even if the latter are spatially separated at arbitrary distances. This is in stark contrast with the predictions of alternative frameworks~\cite{Amico20081,Horodecki:2009zz,Guhne20091,Laflorencie20161} that resort to additional variables, hidden to the observer~\cite{Genovese:2005nw}, to yield  deterministic descriptions based on the classical notions of separability and locality. A decisive criterion for discriminating between local and nonlocal descriptions of quantum phenomena was introduced by J. Bell through the formulation of an inequality~\cite{Bell:1964,bell2004speakable} amenable to experimental verification. The test uses measurements of correlations between non-commuting observables that are independently probed on spatially separated subsystems of an entangled system. The results are then combined into an inequality that any local deterministic theory must satisfy, but which quantum mechanics can, crucially, violate.

Violations of the Bell inequality were first observed in low energy experiments involving photons~\cite{Aspect:1982fx,Weihs:1998gy,Hensen:2015ccp,Giustina:2015yza} and atoms~\cite{Rosenfeld:2017rka}. However, the phenomenology of entanglement and nonlocality can also be investigated in the complementary high-energy regime by using suitable properties of the elementary particles investigated at collider experiments.

Renewed interest in this line of research arose after it was shown that spin correlations in top-quark pairs produced at the CERN Large Hadron Collider (LHC) could exhibit entanglement~\cite{ATLAS:2023fsd,CMS:2024pts,Afik:2020onf} and even nonlocality~\cite{Fabbrichesi:2021npl}. This resurgence has prompted extensive theoretical and phenomenological studies, primarily centered on top-quark production~\cite{Severi:2021cnj,Aguilar-Saavedra:2022uye,Larkoski:2022lmv,Afik:2022kwm,Fabbrichesi:2022ovb,Han:2023fci,Dong:2023xiw,Fabbrichesi:2025psr}, as well as on mesons~\cite{Fabbrichesi:2023idl,Gabrielli:2024kbz,Fabbrichesi:2024rec}, hyperons~\cite{Gong:2021bcp,Fabbrichesi:2024rec}--where nonlocality was shown in experimental data--tau-lepton pairs~\cite{Ehataht:2023zzt,Fabbrichesi:2024wcd}, Higgs boson decays to gauge bosons~\cite{Barr:2021zcp,Ashby-Pickering:2022umy,Aguilar-Saavedra:2022wam,Fabbrichesi:2023cev}, weak diboson production~\cite{Aoude:2023hxv,Fabbrichesi:2023cev}
and vector boson scattering processes~\cite{Morales:2023gow}. For recent reviews on the topic see~\cite{Barr:2024djo,Gabrielli:2025ybg}.

Here, we show the potential of future lepton colliders~\cite{Bernardi:2022hny,
Delahaye:2019omf} to probe quantum entanglement and nonlocality with particularly sensitive processes, such as the tau-lepton, $WW$ and $ZZ$ pair production. Crucially, these channels cannot be tested with comparable precision at hadron colliders, including the LHC and proposed future hadron facilities, for the inherent limit of the experimental platform.

%%%%%%%%%%%%%%%%%%%%%%%%%%%%%%%%%%%%%%%%%%%%%%%%%%%%%%%%%%%%%%%%%%%%%%%%%%%%%
\section{The toolbox}\label{sec1}
We introduce here the main observables that quantify the amount of entanglement and nonlocality in the context of bipartite systems, implemented at a collider experiment by the spin state of created particle pairs. Our discussion is restricted to the cases of spin-1/2 and spin-1, modeled as bipartite qubit and qutrit systems, respectively. These will be described by a density matrix, denoted by $\rho$, which encapsulates the polarizations and spin correlations characterizing the two particles under study.

The spin state of a system made of two spin-$J$ particles is encapsulated in a $(2J+1)^2 \times (2J+1)^2$  density matrix, which can be computed as 

\begin{equation}
\rho_{(\lambda_1, \lambda_1^{\prime})( \lambda_2, \lambda_2^{\prime})} = \frac{M(\lambda_1, \lambda_2) M^{\dag}(\lambda_1^{\prime}, \lambda_2^{\prime})}{|\overline{M}|^2}\,,  
\end{equation}
where $\lambda_i$ and $\lambda_i^\prime$, with $i=1,2$ are the spin label of the $i$-th particle, $M(\lambda_i, \lambda_j)$ is the polarized amplitude for the underlying production process, and $|\overline{M}|^2$ denotes the corresponding spin-summed squared amplitude. All density matrices are positive semi-definite and obey the normalization condition $\Tr(\rho)=1$.

\subsection{Qubits}\label{sec1-1}
For the specific case of a bipartite qubit system, the spin density matrix can be decomposed as
\begin{equation}
\rho
= \frac{1}{4}\Big[ \mathbb{1} \otimes \mathbb{1}
+ \sum_{i=n,r,k} B_i^{\A} (\sigma_i\otimes \mathbb{1})
+ \sum_{j=n,r,k} B_j^{\B}(\mathbb{1} \otimes \sigma_j)
+ \sum_{i,j=n,r,k} C_{ij} (\sigma_i\otimes\sigma_j) \Big],
\label{rhoqubit}
\end{equation}
on the basis formed by the tensor products of the Pauli matrices, $\sigma_i$,  and the $2\times2$ identity matrix $\mathbb{1}$. The coefficients
$B_i^{\A}=\Tr[\rho\,(\sigma_i\otimes\mathbb{1}))]$ and $B_i^{\B}=\Tr[\rho\,(\mathbb{1}\otimes\sigma_i)]$ describe the polarization of the particle $A$ and $B$, respectively. The matrix $C_{ij}=\Tr[\rho\,(\sigma_i\otimes\sigma_j)]$, instead, encodes their spin correlations. All these coefficients can be computed from the amplitude of the underlying production process and depend on the kinematics of the particle pair. Experimentally, they can be measured by reconstructing suitable angular distributions of the $A$ and $B$ decay products. The decomposition refers to a right-handed orthonormal basis, $\{\hn, \hr, \hk\}$, and the spin quantization axis is taken along $\hk$, so that $\sigma_k\equiv\sigma_3$. 

\underline{\bf Entanglement:}\\
The entanglement in a bipartite qubit system can be promptly quantified by the concurrence $0\leq\mathscr{C}\leq 1$. This is analytically evaluated by using the auxiliary matrix
$R=\rho \,  (\sigma_y \otimes \sigma_y) \, \rho^* \, (\sigma_y \otimes \sigma_y)$,
where $\rho^*$ is obtained from $\rho$ upon a complex conjugation of the entries. Although non-Hermitian, the matrix $R$ has non-negative eigenvalues $r^2_i$, $i=1,2,3,4$, which we assume ordered by decreasing value. The concurrence of the state $\rho$ is then given by~\cite{Wootters:PhysRevLett.80.2245}
$\boxed{\mathscr{C} = \max \big( 0, r_1-r_2-r_3-r_4 \big)}$, with a vanishing value signalling the absence of entanglement. 

\underline{\bf Bell inequality:}\\
The Clauser–Horne–Shimony–Holt (CHSH)~\cite{CHSH1969,CHSH1970} formulation of the  Bell inequality for a system formed by two qubits involves the correlation matrix $C$, in \eq{rhoqubit}, and four measurement directions, ${\vec n}_i$, which the two involved observers are free to choose: 
${\vec n}_1\cdot C \cdot \big({\vec n}_2 - {\vec n}_4 \big) +
{\vec n}_3\cdot C \cdot \big({\vec n}_2 + {\vec n}_4 \big)\leq 2$ .
The maximal score can be computed analytically for a general correlation matrix $C$ by considering the two largest eigenvalues $m_1$ and $m_2$ of the $3\times3$ matrix $M = C C^{T}$. Then, if the following (Horodecki condition) holds~\cite{Horodecki:1995nsk}
$
\boxed{
\mathfrak{m}_{12}\equiv m_1 + m_2 >1}\, ,
$
there exist measurement directions ${\vec n}_i$ for which the Bell inequality above is violated with a score of $2\sqrt{\mathfrak{m}_{12}}$. 

\subsection{Qutrits}\label{sec1-2}
The $9\times 9$ density matrix describing the spin state of a pair of qutrits can be decomposed on the basis formed by the tensor products of the Gell-Mann matrices, $T^a$ with $a=1,\dots,8$ and the $3\times3$ unit matrix $\mathbb{1}_3$ as follows:
\bea
\label{eq:rhone}
\rho
= \Big[
  \frac{\mathbb{1}_3\otimes
  \mathbb{1}_3}{9}
    +
    \sum_{a=1}^8 f_a (T^a \otimes \mathbb{1}_3)
    +
    \sum_{a=1}^8 g_a (\mathbb{1}_3\otimes T^a) 
    +\sum_{a,b=1}^8 h_{ab}  (T^a\otimes T^b)\Big]\, .
\label{rho-qutrit}
\eea
An alternative decomposition in terms of spherical tensors is provided in~\cite{Aguilar-Saavedra:2015yza,Bernal:2023jba}. The $f_a$ and $g_a$ coefficients are the vector and tensor polarizations of the two particles, while the $h_{ab}$ matrix encodes their spin correlations.

\underline{\bf Entanglement:}\\
Quantifying entanglement in a bipartite qutrit system is generally a hard numerical task as no analytical expression for any entanglement measure holds in general. Nevertheless, the presence of entanglement can be revealed by investigating the lower bound $\mathscr{C}_2$ of the concurrence, which can be expressed in terms of the coefficients $f_a$, $g_a$, and $h_{ab}$, as 
$\mathscr{C}_2
= 2\max \Big[ -\frac{2}{9}-12 \sum_a f_{a}^{2} +6 \sum_a g_{a}^{2} + 4 \sum_{ab} h_{ab}^{2}
-\frac{2}{9}-12  \sum_a g_{a}^{2} +6 \sum_a f_{a}^{2} + 4 \sum_{ab} h_{ab}^{2},\, 0 \Big]$. 
Throughout this work, a positive value of $\mathscr{C}_2$ serves as a witness of entanglement in the system, without providing a measure of its magnitude.

\underline{\bf Bell inequality:}\\
The Collins-Gisin-Linden-Massar-Popescu (CGLMP) inequality provides a reformulation of the CHSH inequality tailored specifically to a bipartite qutrit system~\cite{Collins:2002sun,Kaszlikowski:PhysRevA.65.032118, Horodecki:1995340,Brunner:RevModPhys.86.419}:
\\
$
\II = P(A_1 = B_1 ) + P(B_1 = A_2 + 1) + P(A_2 = B_2) + P(B_2 = A_1)
-P(A_1 = B_1 - 1) - P(A_1 = B_2)- P(A_2 = B_2 - 1) - P(B_2 = A_1 - 1) \,.
$
The measurements $\hat{A}_i$ and $\hat{B}_j$, with $i, j$ either 1 or 2, have respectively outcomes $A_i$ and $B_j$ which take three possible values: 0,1 and 2. The inequality then uses the probabilities $P (A_i = B_j + k)$ that the outcomes differ by $k$ modulo 3 and, like for CHSH case, deterministic local models inevitably yield $\II \le 2$. The CGLMP inequality can be expressed as the expectation value of a suitable Bell operator $\Bell$, 
$\boxed{\II = \Tr[\rho \Bell]}$~\cite{Collins:2002sun,Latorre:PhysRevA.65.052325}, and the maximal score is obtain upon an optimization procedure involving unitary transformations of the measurement observables, 
$\Bell \to (U \otimes V)^{\dag} \cdot \Bell \cdot (U \otimes V)$, 
where $U$ and $V$ are $3\times3$ unitary matrices operating on each qutrit separately.

\section{Entanglement and nonlocality at lepton colliders}\label{sec3}
Here we present the main results on the Standard Model predictions of quantum entanglement and nonlocality at future lepton colliders. We present the cases of the tau-pair production from electron initial states, as well as of $WW$ and $ZZ$ boson production at electron and muon colliders. Similar results will hold for linear colliders running at the same center-of-mass energies, albeit a possible rescaling of the statistical errors by the characteristic luminosity ratios. 

\subsection{Tau-lepton pair production}\label{sec3-1}
The prospects for observing quantum entanglement and nonlocality with tau pairs have been investigated in~\cite{Ehataht:2023zzt} at the setup of Belle II, with $\sqrt{s}\simeq 10$ GeV, and in~\cite{Fabbrichesi:2024wcd} for the FCC-ee running at the $Z$-pole. 

\begin{figure}[h!]
\begin{center}
\includegraphics[width=2.0in]{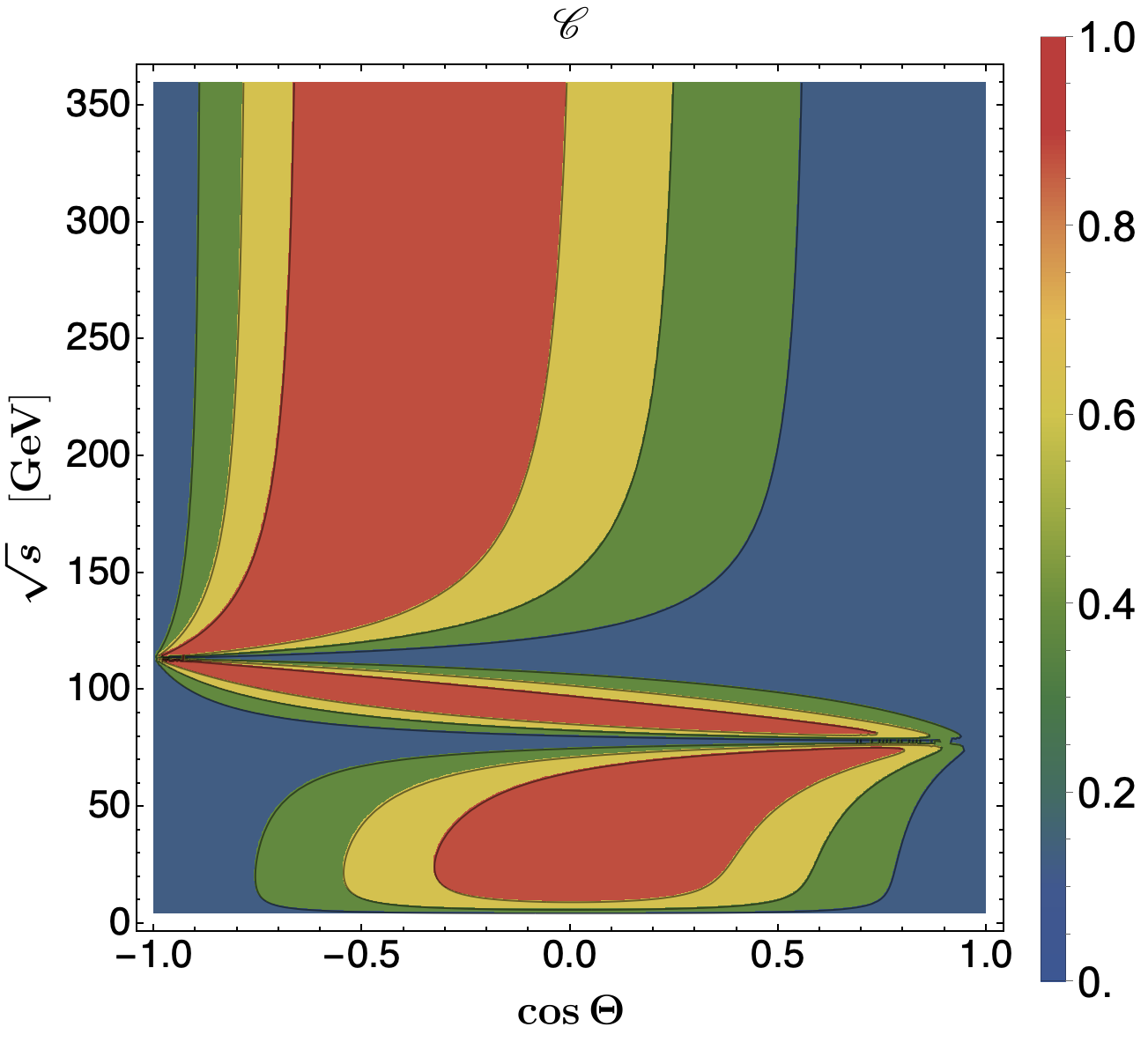}
\hspace{.5cm}
\includegraphics[width=2.0in]{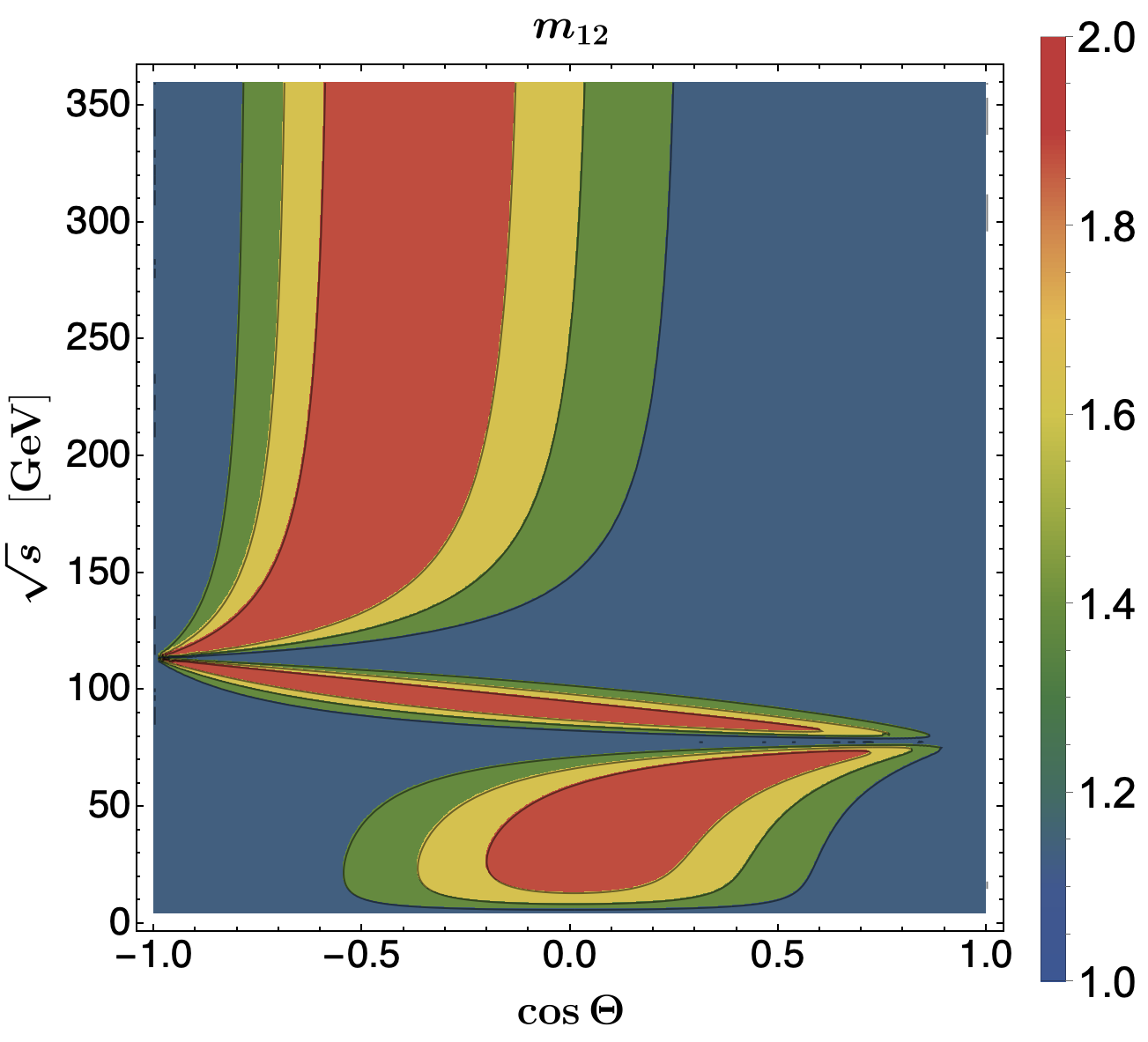}
\caption{\footnotesize  
Standard Model predictions for the entanglement (quantified by the concurrence; left panel) and for nonlocality (right panel) in tau pair spin correlation as a function of the center of mass energy and scattering angle~\cite{Fabbrichesi:2024wcd}. Quantum entanglement is indicated by $\mathcal{C}>0$, while nonlocality is signaled by $\mathfrak{m}_{12}>1$.
\label{fig:ent}  
} 
\end{center}
\end{figure}

The density matrix in \eq{rhoqubit} can be computed from the SM amplitudes of the underlying $e^+e^- \to \tau^+\tau^-$ process. In this case, in the tau pair center of mass frame, our conventions for the $\{\hn,\, \hr,\,\hk\}$ triad in \eq{rhoqubit} set $\hn = \frac{1}{\sin \Theta }\qty(\hp \times\hk), \quad \hr = \frac{1}{\sin \Theta }\qty(\hp-  \hk \cos \Theta)$, where $\hk$ is the direction of the $\tau^+$ momentum and $\Theta$ is the scattering angle. We take $\hp\cdot\hk = \cos\Theta$, with $\hp$ being the direction of the incoming $e^+$. Analytical expressions of the density matrix parameters are reported in~\cite{Ehataht:2023zzt,Fabbrichesi:2024wcd}, while the expected kinematic profiles of the concurrence, $\mathscr{C}$, and the Horodecki parameter, $\mathfrak{m}_{12}$, are shown in Fig.~\ref{fig:ent}. 

%%%%%%%%%%%%%%
%%%%%%%%%%%%%%%%%%%%%%

Alternatively, the coefficients the $B^{\A,\B}_i\equiv B^{\pm}_i$ and $C_{ij}$ can be reconstructed from data (or Monte Carlo simulations) via quantum state tomography, by tracking the angular distributions of suitable $\tau$-pair decay products that act as polarimetric vectors. For instance, a particularly clean reconstruction uses the charged pions emitted in the process
$e^{+}e^{-} \to \tau^{+}\tau^{-} \to \pi ^{+}\pi^{-} \nu_\tau\,\bar \nu_{\tau}$ to give
\begin{equation}
  \frac{1}{\sigma} \,\frac{\dd\sigma}{\dd\cos\theta^\pm_i} = \frac{1}{2}\,
  \qty(1\mp\BB^\pm_i\cos\theta^\pm_i) \,,~~
  \frac{1}{\sigma} \, \frac{\dd\sigma}{\dd\cos\theta^+_{i} \, \dd\cos\theta^-_{j}} = \frac{1}{4} \, \left( 1 + \CC_{ij} \, \cos\theta^+_{i} \, \cos\theta^-_{j} \right) \, ,
  \label{eq:xsec2d}
\nonumber
\end{equation}
in which $\cos\theta^\pm_{i}$ are the projections of the $\pi^\pm$ momentum direction on the $\{\hn, \hr, \hk\}$ basis, as computed in the rest frame of the decaying $\tau^\pm$.

Ref~\cite{Fabbrichesi:2024wcd} presented a full Monte Carlo simulation mimicking the FCC-ee/LEP3 setup, including the effects of initial state radiation (ISR), reconstruction algorithms and the projected detector sensitivity. The implied theoretical uncertainties are very small due to the electroweak nature of the analyzed process. With a benchmark luminosity of 17.6${\rm fb}^{-1}$, it is found for the concurrence
$\mathscr{C}$ and the Horodecki's condition $\mathfrak{m}_{12}$ the following
  results
\be
\boxed{\mathscr{C} = 0.4805 \pm 0.0063|_{\rm stat} \pm 0.0012|_{\rm syst}}\,, 
~~
\boxed{\mathfrak{m}_{12} = 1.239 \pm 0.017|_{\rm stat} \pm 0.008|_{\rm syst}} \,, 
\ee
in which the quoted systematic errors account for ISR and detector effects. As shown by these results, the overall significance for the detection of nonlocality is about 13$\sigma$ (standard deviations) once the errors are added in quadrature, and can reach up to 30$\sigma$ once 150${\rm fb}^{-1}$ of data are collected \cite{Fabbrichesi:2024wcd}.

\subsection{$WW$ and $ZZ$ production}\label{sec3-1}
The prospects for measuring entanglement and nonlocality in bipartite qutrit systems, by using the spin of massive gauge bosons produced at collider experiments, have been analyzed in~\cite{Ashby-Pickering:2022umy,Fabbrichesi:2023cev,Fabbrichesi:2023jep} for the processes
$pp \to W^+ W^-, ZZ, WZ$ studied at the LHC, and 
$\ell^+\ell^-\to W^+W^-,~~ZZ$ at lepton colliders ($\ell=e,\mu$). We focus here on the latter possibility.

The spin density matrix of the diboson system has been derived analytically within the SM framework in \cite{Fabbrichesi:2023cev}. Again, the same density matrix can be extracted by means of quantum state tomography, by analyzing the angular distribution of the decay modes of the final diboson states, in full analogy on what has been done for the tau-pairs. The correlation coefficients $h_{ab}$ and polarization coefficients $f_a$, $g_a$ appearing in the polarization density matrix in \eq{rho-qutrit} can be obtained from the single- and double-differential cross sections via projections. In particular, for the $W^+W^-$ production one has
$
h_{ab} =  \frac{1}{4\, \sigma} \int \int \frac{\di \sigma}{\di\Omega^{+}\,\di\Omega^{-}} \, \mathfrak{p}_+^a \, \mathfrak{p}_-^b \,\di \Omega^+ \di \Omega^-$, and
$
(f_{a}\,,\,g_{a}) =  \frac{1}{2\, \sigma}  \int  \frac{\di \sigma}{\di\Omega^{+}} \, (\mathfrak{p}_+^a\,,\,\mathfrak{p}_-^a), \, \di \Omega^+ $,
where $\sigma$ is the production cross section of the two gauge bosons decaying into the leptonic modes, and $\di \Omega^\pm = \sin \theta^\pm \di \theta^\pm \di \phi^\pm$ are the solid angles of the final-state leptons in the progenitor
$W^{\pm}$ boson rest frames. Here, $\mathfrak{p}_{\pm}^n$ are functions depending on the $\Omega^{\pm}$ solid angles, whose expressions can be found in \cite{Barr:2024djo,Barr:2021zcp}.

%%%%%%%%%%%%%%
\begin{figure}[h!]
\begin{center}
\includegraphics[width=2.0in]{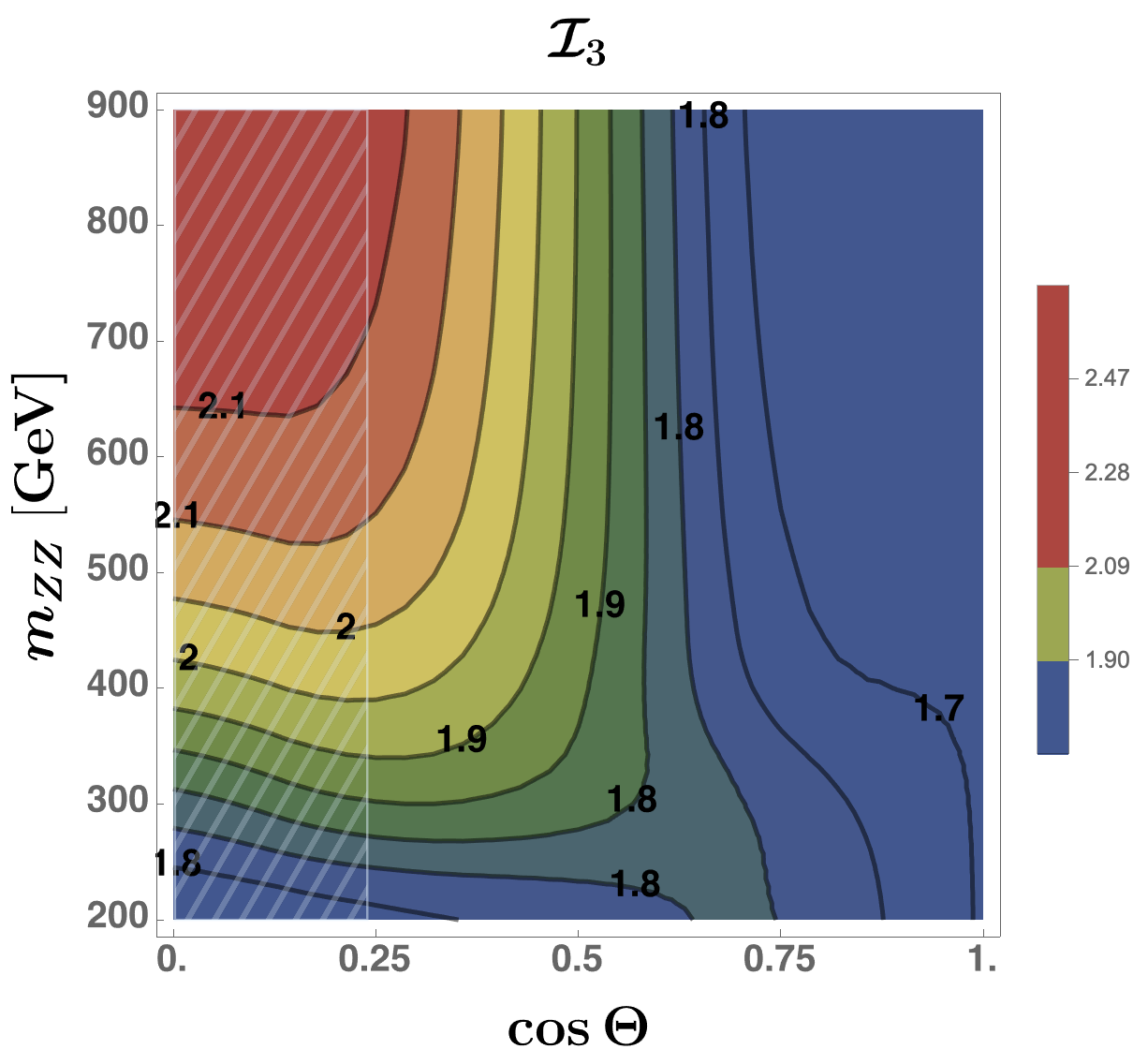}
\hspace{.5cm}
\includegraphics[width=2.0in]{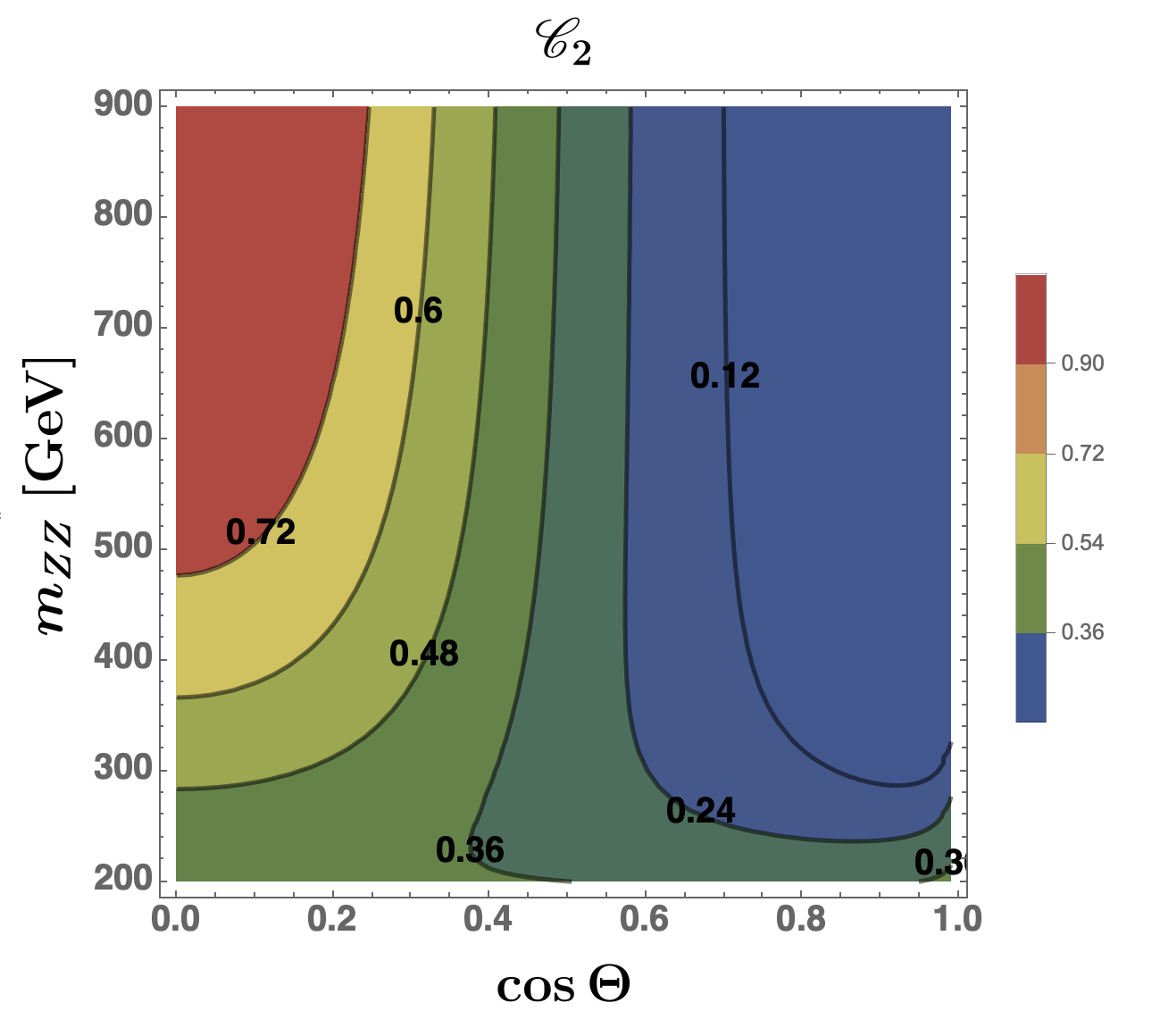}
\caption{\small The observables ${\cal I}_{3}$ and $\cmb$ for the process $\ell^+\ell^-\to Z Z$ as functions of the diboson invariant mass and scattering angle in the center of mass frame. The hatched area in the plot on the left represents the bin in which nonlocal spin correlations are observable.
\label{fig:ZZmuon} 
}
\end{center}
\end{figure}
 %%%%%%%%%%%%%%%%%%%%%%

In Fig.\ref{fig:ZZmuon} we show the results pertaining to the $ZZ$ production ~\cite{Fabbrichesi:2023cev} within the SM. The plots show the predictions of the lower bound on concurrence, $\cmb$, and on nonlocality, encapsulated in ${\cal I}_3$, as a function of kinematic parameters. Analogous results for the  $WW$ mode can be found in~\cite{Fabbrichesi:2023cev}, and show that nonlocality can be probed at a lepton collider across a broader region of the relevant kinematic space than at the LHC, with a greater magnitude and minimal theoretical uncertainties. In contrast, for the $ZZ$ case, the kinematic domain in which nonlocality is found is comparable to that accessible at the LHC. These studies relied on leading-order cross sections computed with the \texttt{MadGraph5\_aMC@NLO} generator~\cite{Alwall:2014hca}, with event yields rescaled by the lepton identification efficiency. Following a conservative assumption consistent with the LHC analyses, this efficiency was taken to be $70\%$ per lepton.

With benchmark center of mass energies of $\sqrt{s}=1~\mathrm{TeV}$, achievable by a muon collider~\cite{Delahaye:2019omf}, and $\sqrt{s}=368~\mathrm{GeV}$ of a future lepton collider~\cite{Bernardi:2022hny}, Ref.~\cite{Fabbrichesi:2023cev} showed that the null hypothesis ${\cal I}_3 \leq 2$ can be rejected at a statistical significance of $2\sigma$ with the $WW$ mode in both the setups. A similar results holds for the $ZZ$ production at a muon collider, whereas the same channel can reach a significance of $4\sigma$ at future electron machines due to the higher achievable event rate~\cite{Fabbrichesi:2023cev}.

\section{Conclusions}
The results discussed show that future lepton colliders offer a unique opportunity to probe quantum entanglement and nonlocality in a high-energy environment complementary to the setups traditionally used in quantum mechanics experiments. The study focused on the case of bipartite qubit and qutrit systems, implemented by tau-lepton and weak gauge boson pairs, respectively. The clean experimental environment and very high luminosity enable measurements with unprecedented sensitivity, well beyond the reach of present and future hadron colliders. Lepton colliders therefore offer the most promising avenue for exploring the phenomenology of quantum correlations in the electroweak sector at high energies.

\section*{Acknowledgements}
This work was supported by the Estonian Research Council grants TARISTU24-TK10, TARISTU24-TK3, RVTT3. PRG1884, and by the CoE TK 202 ``Foundations of the Universe’'.
\bibliography{proceedings-LCWS}

\end{document}